\documentclass[12pt]{iopart}
\newcommand{\lambdav}{\boldsymbol\lambda}

\newcommand{\mut}{\tilde{\bmu}}

\newcommand{\Ut}{\tilde{U}}
\newcommand{\Deltat}{\tilde{\Delta}}

\newcommand{\sgn}{\textrm{sgn}}
\newcommand{\Gt}{\tilde{G}}
\newcommand{\Sigmat}{\tilde{\Sigma}}
\newcommand{\Gammat}{\tilde{\Gamma}}
\newcommand{\Ueff}{\tilde{U_e}}
\newcommand{\Pit}{\tilde{\Pi}_{ph}}

\usepackage{iopams, graphicx}  

\begin{document}

\title{Renormalized perturbation theory flow equations for the Anderson impurity model}
\author{V Pandis}

\address{Department of Mathematics, Imperial College London, London SW7 2AZ, United Kingdom}
\ead{vassilis.pandis@gmail.com}

\begin{abstract}
We apply the renormalized perturbation theory (RPT) to the symmetric Anderson impurity model. Within
the RPT framework exact results for physical observables such as the spin and charge 
susceptibility can be obtained in terms of the renormalized values $\mut = (\Deltat, \Ut)$ of the hybridization $\Delta$ and Coulomb interaction $U$ of the model. 
The main difficulty in the RPT approach 
usually lies in the calculation of the renormalized values themselves. In the present work we show how this
can be accomplished by deriving differential flow equations describing the evolution of $\mut(\Delta)$
with $\Delta$. By exploiting the fact that $\mut(\Delta)$ can be determined analytically in the limit $\Delta \rightarrow \infty$ we solve the flow 
equations numerically to obtain estimates for the renormalized parameters
in the range $0<U/\pi\Delta<3.5$.
\end{abstract}
\maketitle

\section{Introduction}

The Anderson impurity model (AIM)~\cite{Anderson61} was introduced more than five decades ago in an effort to explain the
presence of localized magnetic moments in metals. Since then it has been the subject of intensive theoretical 
study and established itself as a standard test-bed for methods in strong correlation physics. The AIM has also enabled
significant progress in the field of lattice impurity models through the Dynamical Mean Field Theory~\cite{Georges96}, which allows such models to be
mapped onto an auxilliary single-impurity problem. Finally, it has recently enjoyed renewed popularity as a 
model for quantum dots~\cite{Hershfield91, Landauer92}.  

A review of the literature on the AIM is well beyond the scope of this article; a comprehensive account can be found in \cite{Hewson97}.
Early insight into the model was provided by the work of Yosida and Yamada~\cite{Yosida70, Yamada75}, who calculated properties of the
model using a perturbation expansion in the Coulomb repulsion $U$. While this approach is asymptotically exact 
in the limit $U\rightarrow 0$, it is unable to describe the model in the physically interesting regime
of large $U$. Subsequently, insight into this regime was gained through the application of non-perturbative 
methods such as the Numerical Renormalization Group~\cite{Wilson75, Krishna-murthy80} and the Bethe Ansatz~\cite{Andrei83, Wiegmann83}. These in turn have led to an
appreciation of the role that low-energy excitations play in the emergence of strong correlation effects
such as the appearance of the Kondo scale.

In this article we will focus on the Renormalized Perturbation Theory (RPT) approach to the AIM~\cite{Hewson93b, Hewson01}. 
In this approach one replaces the original constants of the model with renormalized values and 
introduces counter-terms to avoid overcounting. This can also be thought of as a reorganization of
the (bare) perturbation expansion. The advantage of this approach is that it leads to $\emph{exact}$
predictions for the magnetization, the charge and spin susceptibilities and the $\omega^2$ coefficient
of the self-energy. 

The main challenge in the RPT lies with the determination of the renormalized parameters
themselves. Typically, this has been done with the help of some external non-perturbative method, notably the
NRG~\cite{Hewson04}. Recently however a method based on flow equations has been developed~\cite{Edwards11b, Pandis13} which allows their determination
exclusively within the RPT framework, obviating the need for the NRG. To implement the flow equation
approach one first identifies the parameter $\alpha$ of the original model that will be varied. The
second stage involves deriving a differential equation for $d\mut/d\alpha$. Finally, one has
to determine a limit $\alpha\rightarrow\alpha_0$ in which $\mut(\alpha)$ can be analytically
determined and use these values as a boundary condition to the system of the differential equations.
By integrating the system from $\alpha_0$ to more realistic values of $\alpha$ one can thus
determine the renormalized parameters for a physically-relevant AIM. 

This flow equation approach was introduced in \cite{Edwards11b, Pandis13} which examined the AIM at half-filling
by introducing a magnetic field $h$ and exploiting the fact that the spin fluctuations from which
the strong correlation effects stem are frozen out when $h\rightarrow\infty$, a limit in which the AIM
can be solved using mean field theory. In \cite{Pandis13} the flow equation programme
was also carried out for the asymmetric model by varying $\epsilon_d$ and exploiting again the
applicability of the mean field in the $\epsilon_d\rightarrow -\infty$ limit. 

In the present article we consider the symmetric Anderson model, confining ourselves for simplicity to the case of zero
magnetic field, and examine the effects of varying the hybridization $\Delta$. We note that since the
dynamics of the model are governed by the ratio $u = U/\pi\Delta$, taking $\Delta \rightarrow \infty$ 
is sufficient to put us into the tractable weak correlation regime. Compared to varying $U$ directly
this approach has the advantage that the model remains symmetric throughout the renormalization
procedure allowing us to carry out the flow without having to vary $\epsilon_d$ and $U$ simultaneously.
Furthermore, since the renormalized level $\tilde{\epsilon_d}$ parameter is always equal to zero for
a symmetric AIM with no magnetic field, we only have two, rather than three, flow equations to
concern ourselves with.

\section{Renormalized Perturbation Theory}

The Anderson model in the limit of a wide flat band can be studied in the functional formalism by means of the effective Lagrangian
\begin{equation}
\mathcal{L} (\Delta, U) = \sum_{\sigma = \uparrow, \downarrow} \overline{d}_\sigma(\tau) \left(\partial_\tau -U/2 - \rmi\Delta\right) d_\sigma(\tau)
+ U n_{\uparrow}(\tau) n_{\downarrow}(\tau),
\label{eq:lagbare}
\end{equation}
where $n_\sigma(\tau) = \overline{d}(\tau)d(\tau)$ and $\overline{d}(\tau),d(\tau)$ are the usual Grassmann-valued fields.
To set up the bare perturbation theory we have to separate $\mathcal{L}$ into a non-interacting part $\mathcal{L}_0$ and an interacting part $\mathcal{L}_I$
which we choose as follows:
\begin{eqnarray}
\mathcal{L}_0 (\Delta) =  \sum_{\sigma = \uparrow, \downarrow} \overline{d}_\sigma(\tau) \left(\partial_\tau - \rmi\Delta\right) d_\sigma(\tau) \nonumber \\
\mathcal{L}_I (U) =  U (n_{\uparrow}(\tau) - 1/2) (n_{\downarrow}(\tau) -1/2).
\end{eqnarray}
The component $\mathcal{L}_0$ gives rise to a non-interacting Green's function 
\begin{equation}
G^{(0)} (\omega) = \frac{1}{\omega +\rmi\Delta \sgn(\omega)}
\end{equation}
which is the starting point of the diagrammatic expansion. By treating $\mathcal{L}_I (U)$ as a perturbation 
and associating internal lines with $G^{(0)}(\omega)$ we can calculate the self-energy $\Sigma(\omega)$ and obtain
the interacting Green's function as usual through the Dyson equation:
\begin{equation}
G(\omega) = \frac{1}{\omega + i\Delta\sgn(\omega) - \Sigma(\omega)}.
\label{eq:G}
\end{equation}

In the renormalized theory one starts by separating the Lagrangian in \eref{eq:lagbare} into a renormalized Lagrangian and a 
counter-term Lagrangian of the same form $\mathcal{L}(\Delta, U ) = \tilde{\mathcal{L}}(\mut) 
+ \tilde{\mathcal{L}_{ct}}(\lambdav(\mut))$ where
\begin{equation}
\tilde{\mathcal{L}}_{ct}(\lambdav) =  \sum_{\sigma = \uparrow, \downarrow}\tilde{\overline{d}}_\sigma(\tau) (\lambda_{2, \sigma} \partial_\tau + \lambda_{1, \sigma})\tilde{d}_\sigma(\tau) + \lambda_3 \tilde{n}_\uparrow(\tau) \tilde{n}_\downarrow(\tau)
\label{eq:renormstd}
\end{equation}
depends on $\mut$ only implicitly through the counter-terms $\lambdav = (\lambda_1, \lambda_2, \lambda_3)$.
To set up the the perturbation theory we separate the Lagrangian $\mathcal{L}$ into renormalized non-interacting and interacting components
\begin{eqnarray}
\tilde{\mathcal{L}}_0 &=  \sum_{\sigma = \uparrow, \downarrow}\tilde{\overline{d}}_\sigma(\tau) \left(\partial_\tau - \rmi\Deltat\right) \tilde{d}_\sigma(\tau) \nonumber \\
\tilde{\mathcal{L}}_I &= \Ut  \tilde{n}_\uparrow(\tau) \tilde{n}_\downarrow(\tau) + \tilde{\mathcal{L}}_{ct}.
\end{eqnarray}
The non-interacting component now gives rise to a renormalized non-interacting Green's function
\begin{equation}
\tilde{G}^{(0)} (\omega) = \frac{1}{\omega +\rmi\Deltat \sgn(\omega)}.
\end{equation}
Note that in addition to the Coulomb interaction, in the renormalized theory the diagrammatics must take into account the interaction terms
provided by the counter-terms. These will give rise to a renormalized self-energy\footnote{We clarify that $\Sigmat(\omega;\mut)$ also depends on 
the counter-terms $\lambdav(\mut)$ but in the interests of notational brevity this is not expressly indicated.}
 $\tilde{\Sigma}(\omega; \mut)$ and an interacting Green's function
$\tilde{G}(\omega)$ defined through the Dyson equation $[\tilde{G}(\omega)]^{-1} = [\tilde{G}^{(0)}(\omega)]^{-1} - \tilde{\Sigma}(\omega;\mut)$.

To determine the counter-terms  as functions of the renormalized parameters we impose the
renormalization conditions
\begin{eqnarray}
\Sigmat (0; \mut) &= 0 \nonumber \\
\Sigmat'(0; \mut) &= 0 \nonumber \\
\Gammat_{\uparrow \downarrow}(0,0,0; \mut)  &= \Ut,
\label{eq:RGeq}
\end{eqnarray}
where $\Gammat_{\uparrow \downarrow} (\omega_1, \omega_2; \omega_3)$ is the reducible four-vertex. Note that
the bare and renormalized Green's functions can be directly related by expanding  $\Sigma(\omega)$ in \eref{eq:G} 
to  $\mathcal{O}(\omega^2)$ and defining the quasi-particle weight as $z = (1 - \Sigma'(0))^{-1}$. We find
then that 
\begin{eqnarray}
\Deltat &= z\Delta \nonumber \\
\Sigmat(\omega; \mut) &= z (\Sigma(\omega) - \omega\Sigma'(0) - \Sigma(0)) \nonumber \\
G(\omega) &= z \Gt(\omega).
\end{eqnarray}

\section{Flow equations}

We now focus on deriving the flow equations by examining the dependence of the renormalized parameters on $\Delta$. Consider the bare Lagrangian of \eref{eq:lagbare} for a model with parameter $\Delta + \delta$. 
This can be rewritten in terms of the renormalized parameters $\mut(\Delta + \delta)$ in the usual manner:
\begin{equation}
\mathcal{L} (\Delta + \delta) = \tilde{\mathcal{L}}_0( \mut(\Delta + \delta)) + \tilde{\mathcal{L}_I}( \mut(\Delta + \delta).
\end{equation}
In this expression  $\tilde{\mathcal{L}}_I$ denotes the interacting part of the Lagrangian and contains the Coulomb and
counter-term vertices.  Note that the Lagrangian of the bare model in \eref{eq:lagbare} is linear in $\Delta$. We can exploit this to separate it into a finite and infinitesimal term 
\begin{equation}
\mathcal{L} (\Delta + \delta) = \mathcal{L}(\Delta) + \mathcal{L}_r(\delta),
\end{equation}
and proceed to renormalize $\mathcal{L}(\Delta)$ as usual, i.e\ in terms of $\mut(\Deltat)$
\begin{equation}
\mathcal{L} (\Delta + \delta) = \tilde{\mathcal{L}}_0( \mut(\Delta)) + \left[ \tilde{\mathcal{L}}_I( \mut(\Delta)  + \tilde{\mathcal{L}}_r(\delta) \right].
\label{eq:secondrenormalization}
\end{equation}
The right hand side of \eref{eq:secondrenormalization} is similar to $\tilde{\mathcal{L}} (\mut)$, save for the appearance of the additional infinitesimal vertex.  
We can exploit the dual representation of the Lagrangian to express the bare Green's function in two ways 
\begin{eqnarray}
G(\omega) &= \frac{z(\Delta + \delta)}{\omega  + \rmi\Deltat(\Delta + \delta) - \Sigmat(\omega; \mut(\Delta + \delta))} \nonumber \\
&= \frac{z(\Delta)}{\omega - + \rmi\Deltat(\Delta) - \Sigmat^{(2)}(\omega; \mut(\Delta))}.
\label{eq:greensdual}
\end{eqnarray}
Here the quantity $\Sigmat(\omega; \mut(\Delta + \delta))$ denotes the conventional renormalized self-energy, obtainable by 
taking into account the renormalized Coulomb and counterterm terms in $\tilde{\mathcal{L}}_I$. The self-energy $\Sigmat^{(2)}(\omega; \mut(\Delta), \delta)$
derives from the bracketed term in \eref{eq:secondrenormalization}. Note that while $\Sigmat(\omega; \mut(\Delta + \delta))$ (and the corresponding four-vertex)
is going to satisfy the renormalization conditions of \eref{eq:RGeq} by definition, this will not be true for $\Sigmat^{(2)}(\omega; \mut(\Delta), \delta)$ as
the presence of the additional vertex will render the counter-term cancellation incomplete. We can extract a formal expression for the
flow equations by equating the two Green's functions of \eref{eq:greensdual} and their frequency derivatives at $\omega=0$. We find that
$\Deltat(\Delta + \delta) = \overline{z}(\Delta; \delta) \Deltat(\Delta)$, where
\begin{equation}
\overline{z}(\Delta; \delta) = \frac{z(\Delta+\delta)}{z(\Delta)}= \frac{1}{1 - \Sigmat^{(2)}(\omega; \mut(\Delta), \delta)}.
\label{eq:Dflow}
\end{equation}
The corresponding differential equation can be deduced by writing $\Sigmat^{(2)}(\omega; \mut(\Delta), \delta) = q(\mut(\Delta)) \delta + \mathcal{O}(\delta^2)$ and taking the limit 
$\delta \rightarrow0$. We thus find that
\begin{equation}
\frac{\partial \Deltat}{\partial \Delta} = q(\mut(\Delta))\Deltat + z(\Delta).
\label{eq:flowdelta}
\end{equation}

We now seek to derive a corresponding flow equation for $\tilde{U}$. In \cite{Edwards11b, Pandis13} this was followed from the Ward identities
which relate $\Ut$ to the derivative of the self-energy with respect to $h$ or $\epsilon_d$. Unfortunately we are not aware of a corresponding
identity with respect to $\Delta$ so we must turn our attention to the diagrammatic representation of the four-vertex.

\begin{figure}
\begin{center}
\includegraphics[scale=0.3]{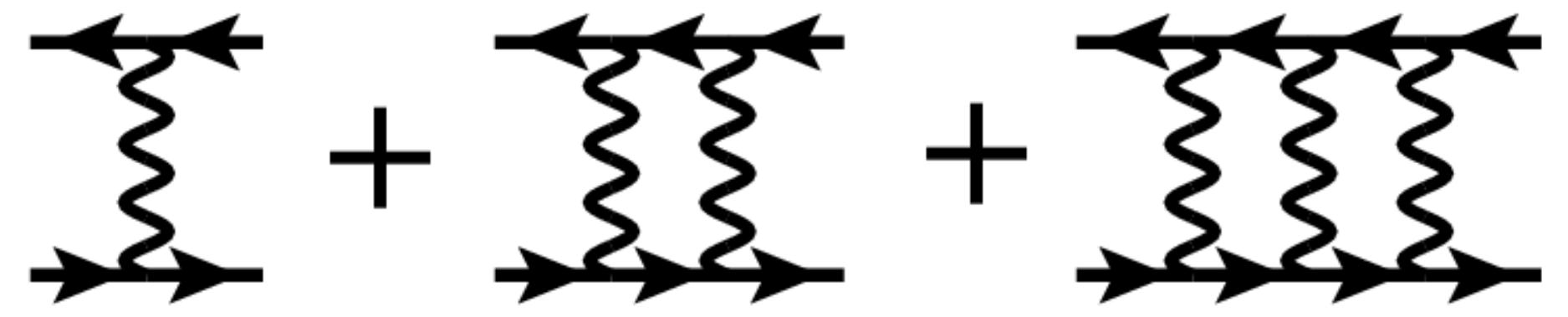}
\end{center}
\caption{The diagrammatic representation of the four-vertex in the particle-hole channel in the RPA approximation. Here the interaction vertices represent the combined interaction vertex
$\Ut + \lambda_3$.}
\label{fig:ph-ladder}
\end{figure}

\begin{figure}
\begin{center}
\includegraphics[scale=0.3]{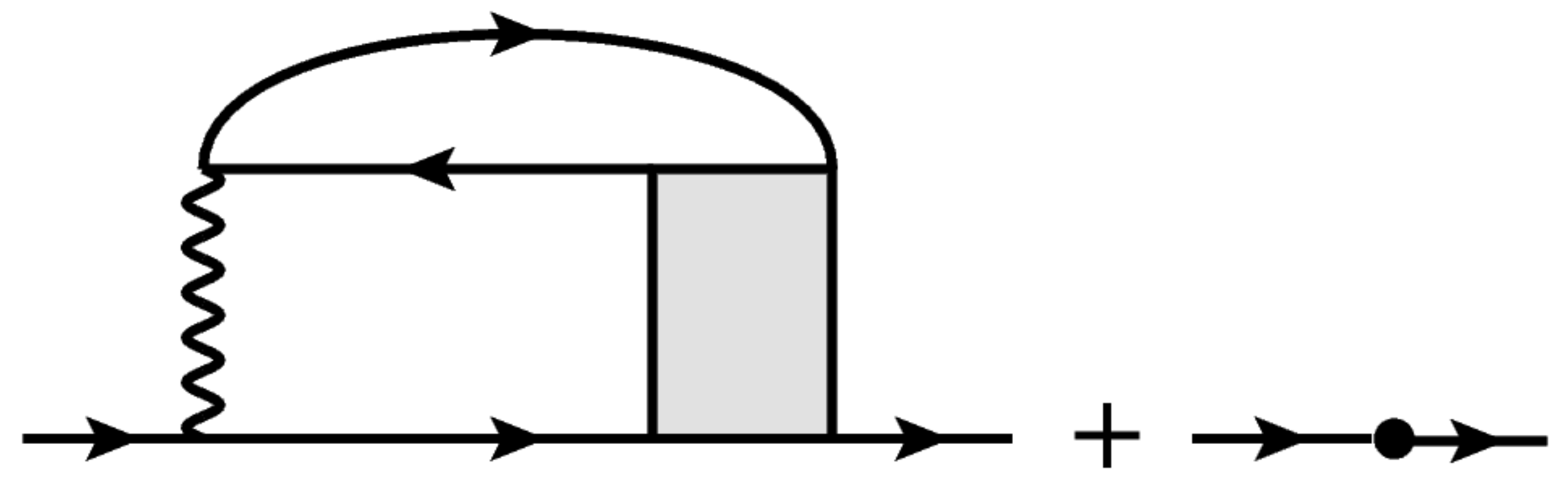}
\end{center}
\caption{Diagrammatic representation of the RRPA in terms of the four-vertex and the tree-level counter-term. }
\label{fig:rpa}
\end{figure}

To constuct an approximation to $\Gammat$ and $\Sigmat$ we resort to the Renormalized Random Phase Approximation (RRPA)~\cite{Hewson06, Hewson06b, Hewson06c}  which is thought to accurately describe the spin-flip excitations principally
responsible for the renormalization effects at half-filling. The four-vertex is constructed by combining the two-body terms into an effective vertex
$\Ueff = \Ut + \lambda_3$ and resumming the particle-hole ladder of Figure~\ref{fig:ph-ladder}. We thus find that 
\begin{equation}
\Ut = \frac{\Ueff}{1 - \Ueff\Pit(0)},
\end{equation}
where $\Pit(\omega)$ denotes the renormalized particle-hole propagator defined as 
\begin{equation}
\Pit(\omega) = \rmi \int_{-\infty}^{\infty} \frac{\rmd \omega'}{2 \pi} \Gt^{(0)}(\omega')\Gt^{(0)}(\omega' + \omega).
\end{equation}
The self-energy is then constructed according to Figure~\ref{fig:rpa} and is equal to 
\begin{equation}
\Sigmat(\omega) = -\rmi\int_{-\infty}^{\infty} \frac{\rmd \omega'}{2 \pi} \Gt^{(0)}(\omega' + \omega) \frac{\Ueff^2 \Pit(\omega')}{1 - \Ueff\Pit(\omega')} + \omega\lambda_2 + \lambda_1.
\end{equation}

We will account for the dependence of $\Ut$ on $\Delta$ by considering the insertion of the infinitesimal vertex into the pair propagator
and, in what constitutes only an approximation,  neglecting the explicit dependence of $\Ueff$ on $\Delta$. We obtain then that
\begin{equation}
\frac{\partial \Ut}{\partial \Delta} = \frac{\Ueff^2/2\pi\Deltat^2}{\left( 1- \Pit(0) \Ueff\right)^2}.
\label{eq:Uflow} 
\end{equation}
Equations \eref{eq:Dflow} and \eref{eq:Uflow} constitute a closed system of differential equations which we will now
supplmenet with boundary conditions.

\section{Boundary conditions and results}

\begin{figure}
\begin{center}
\includegraphics[scale=0.3]{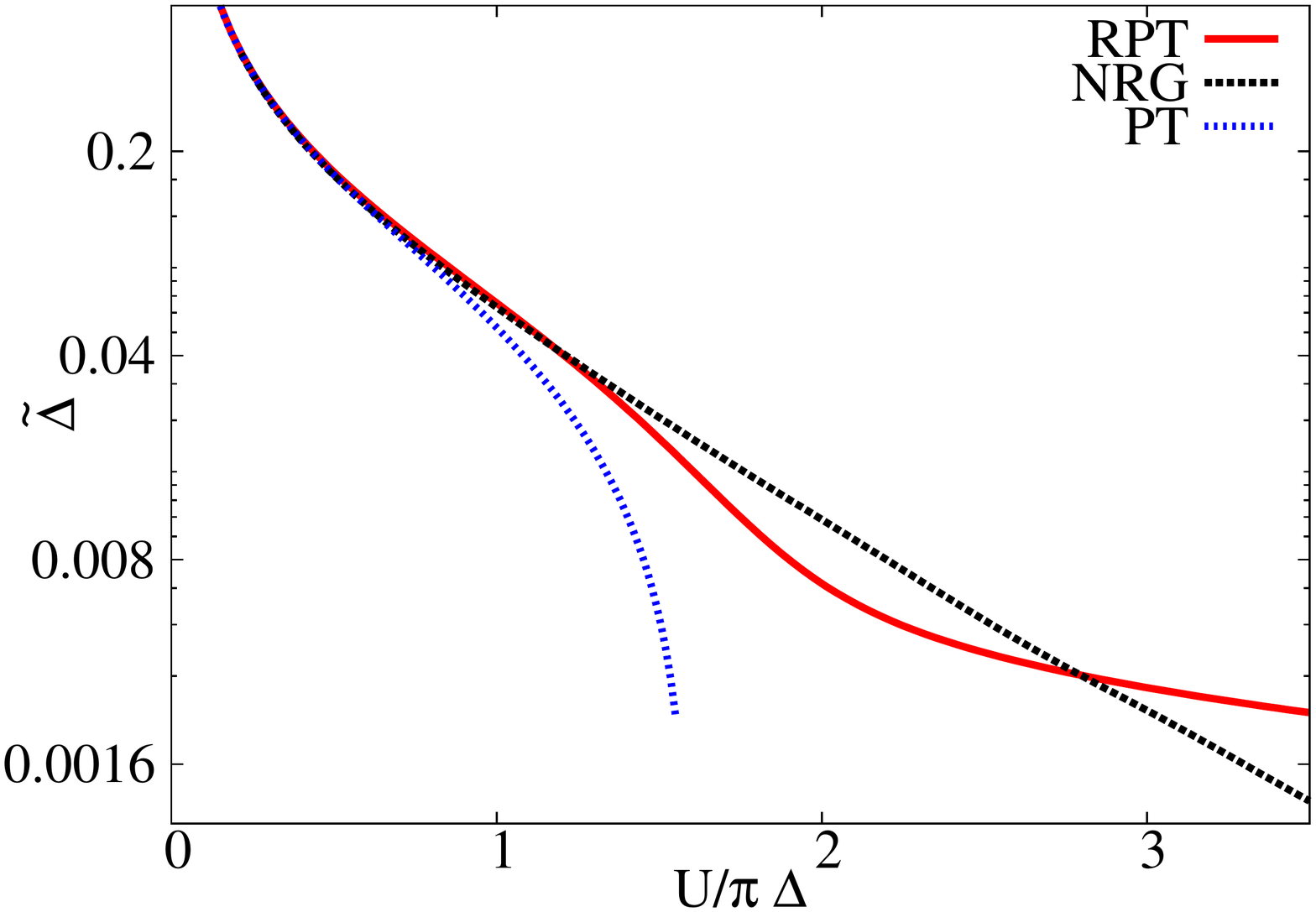}
\end{center}
\caption{The renormalized hybridization  $\Deltat$ is shown as a function of $u = U/\pi\Delta$.}
\label{fig:Dresults}
\end{figure}

\begin{figure}
\begin{center}
\includegraphics[scale=0.3]{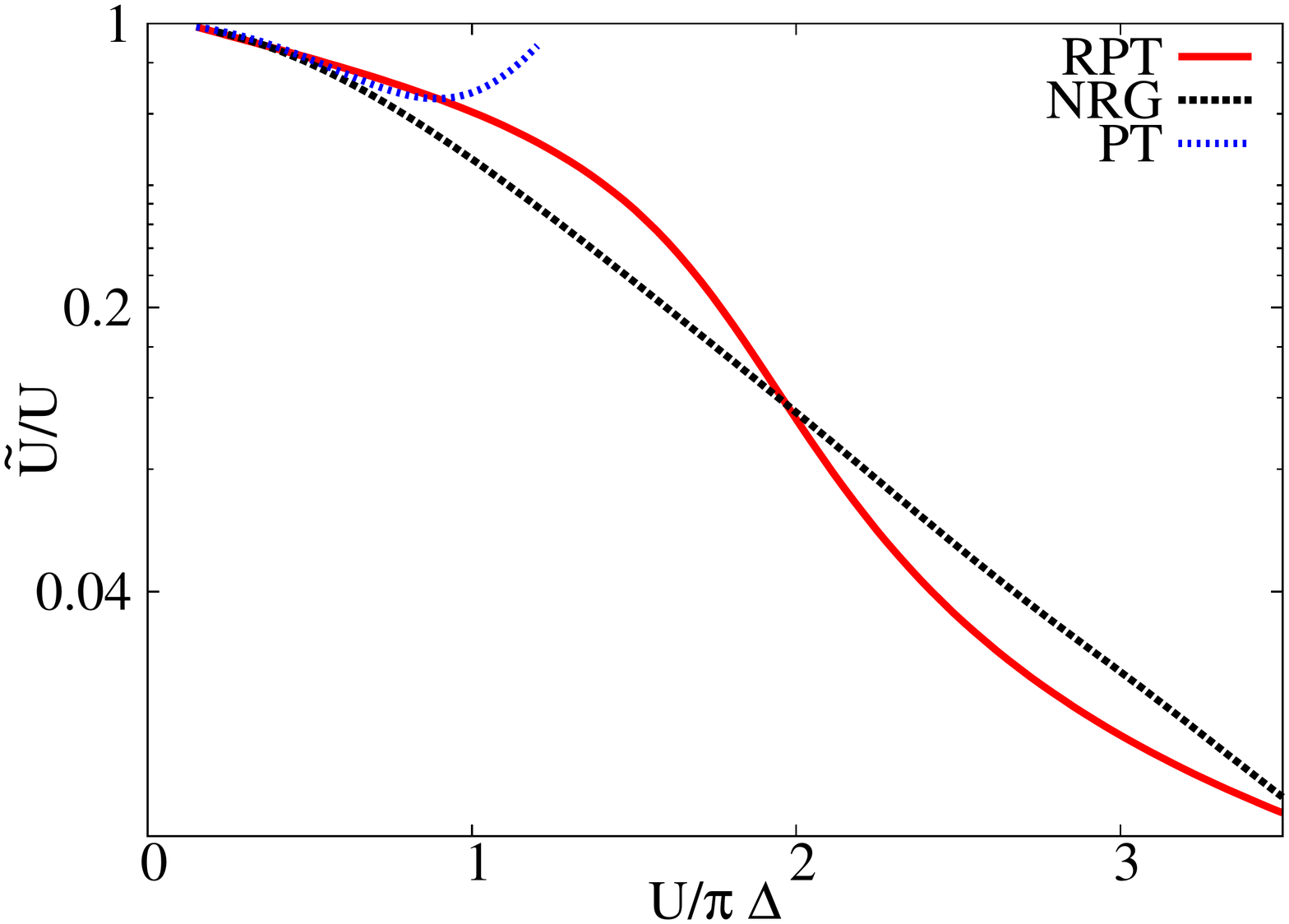}
\end{center}
\caption{The ratio $\Ut/U$ shown as a function of $u = U/\pi\Delta$.}
\label{fig:Uresults}
\end{figure}

In the weak-correlation regime $\Delta \rightarrow \infty$ the results of bare perturbation theory~\cite{Yosida70, Yamada75} can be used to derive~\cite{Hewson01}
the leading terms in the expressions of the renormalized parameters as functions of the bare parameters\footnote{
Noting that $u = \tilde{u} + \mathcal{O}(\tilde{u}^2)$ we could in principle also use these expressions to derive
the flow equations. However, the series in \eref{eq:weakrpt} are truncated to $\tilde{u}^2$ so the validity of any equations derived therefrom would be
similarly restricted to the weak-correlation regime.}. Using the results of \cite{Zlatic83}, which are based on the Bethe Ansatz,  we can in fact deduce all terms
order-by-order but we will not make use of this result here. Working to next-to-next-to-leading order we have that
\begin{eqnarray}
\Deltat &= \Delta \left[ 1 - \left(3-\frac{\pi^2}{4}\right)u^2 + \left(105-\frac{45\pi^2}{4} + \frac{\pi^4}{16}\right)u^4 + \ldots \right] \nonumber  \\
\Ut &= U \left[1 - \left(\pi^2 - 9\right)u^2 + \left(672-\frac{141\pi^2}{2} + \frac{\pi^4}{4}\right)u^4 + \ldots \right].
\label{eq:weakrpt}
\end{eqnarray}
These expressions, truncated to $\mathcal{O}(u^2)$ in the interests of minimalism, will be used as boundary conditions to solve the system 
defined by \eref{eq:Dflow}, \eref{eq:Uflow} numerically using the implementation of the Runge-Kutta-Fehlberg algorithm in the \verb|GSL| 
library~\cite{GSL}. Our results are $\Deltat(\Delta)$, $\Ut(\Delta)$ are shown in Figure~\ref{fig:Dresults} and \ref{fig:Uresults} respectively,
starting from a value of $\Delta$ such that $u=0.15$. As a check we also show the renormalized parameters as determined from the 
NRG following the method in \cite{Hewson04}. Additionally, we have also plotted \eref{eq:weakrpt} to demonstrate that the flow procedure is 
indeed superior to the simple inversion of the results of the bare perturbation theory.

In \ref{fig:Dresults} we observe that
NRG results for $\Deltat$ are reproduced essentially exactly in weak and intermediate correlation regime. Our results start deviating from the
NRG values in the neighbourhood of $u=2$ though the deviation is not amplified as we cross over to the regime of very strong correlations. Note that while in relative
terms the discrepancy from the NRG is not small when $u=3.5$, the value of $\Deltat$ has already reduced by almost three orders of magnitude.
In fact, $\Ut/\pi\Deltat \approx 1$, indicating that we have already entered the Kondo regime where the universal scale $T_K$ emerges.

We now turn our attention to the results for $\Ut/U$ shown in \ref{fig:Uresults}. The results in the weakly correlated regime again agree closely
with the NRG though discrepancies arise as $u$ is increased sooner than in the case of $\Deltat$. For $u\approx1.5$ we observe a significant
departure from the results of the NRG which we attribute to the somewhat crude approximation made in \eref{eq:Uflow} of neglecting the dependence 
of $\Ueff$ on $\Delta$. As $u$ is increased however we find that the slope of the NRG results is reproduced and that rough agreement is maintained
even when $u=3.5$. 

We would like to emphasize that for both parameters the renormalization effects are very strong, with $z \approx 0.09$ when $u=3.5$. We thus consider it a 
success that these can be even approximately described by the flow equation method, despite the fact that a precise
agreement with the results of the NRG has not been achieved. We anticipate that the accuracy of our approach can 
be improved by incorporating the additional corrections due to particle-particle and longitudinal scattering.



\section{Summary}

In this article we have shown how the renormalized theory for the symmetric AIM can be used to estimate the renormalized
parameters by considering infinitesimal changes to the model's hybridization. We described how flow equations can
be obtained by expressing the Lagrangian of the same bare model in terms of renormalized parameters corresponding
to different $\Delta$. We also showed how an approximate flow equation for $\Ut$ can be derived even in the absence of an exact Ward identity.
This led to a closed system of differential equations which we then supplemented with boundary conditions derived from
the results of the original perturbation theory. This was solved numerically and was found to be able to provide an estimate
for the renormalized parameters in the range $u\leq 3.5$. Though our approach does not precisely reproduce the 
parameters it does nonetheless capture the bulk of the renormalization effects even in the strong correlation regime. 
By rendering the RPT independent of the NRG and self-contained we have made progress towards
extending the RPT's utility to problems that cannot be tackled by other non-perturbative methods.

\ack
This work was supported financially by the Engineering and Physical Sciences Research Council. The author would like to 
thank Alex Hewson for many helpful discussions and critical feedback. 

\section*{References}
\bibliographystyle{unsrt}
\bibliography{bibliography}
\end{document}